\documentclass[12pt]{article}
\usepackage{amsfonts}
\usepackage{amsmath}
\usepackage{amssymb}
\usepackage{graphicx}
\usepackage{color}
\usepackage{cite}

\def \be {\begin{equation}}
\def \ee {\end{equation}}
\def \bea {\begin{eqnarray}}
\def \eea {\end{eqnarray}}
\def \nn {\nonumber}
\def \la {\langle}
\def \ra {\rangle}
\def \rr {\raise.35ex\hbox{\small $\prime$}\kern-.17em{\mbox{\large $\imath$}}}
\def \del {\partial}
\def \dels {\partial\kern-.5em / \kern.5em}
\def \As {{A\kern-.5em / \kern.5em}}
\def \Ds {D\kern-.7em / \kern.5em}

\def \a {\alpha}

\def \d {\delta}
\def \eps {\epsilon}

\def \lam {\lambda}
\def \Lam {\Lambda}

\def \II {I\hspace{-.1em}I\hspace{.1em}}

\def \IIA {\mbox{\II A\hspace{.2em}}}

\def \mh {\underline{\mu}}
\def \nh {\underline{\nu}}

\def \Xh {\hat{X}}
\def \Psih {\hat{\Psi}}
\def \Ah {\hat{A}}

\def \Dh {\hat{D}}
\def \Fh {\hat{F}}

\newcommand{\cN}{\mathcal{N}}
\newcommand{\cM}{\mathcal{M}}

\newcommand{\ba}{\begin{eqnarray}}
\newcommand{\ea}{\end{eqnarray}}

\newcommand{\cD}{\mathcal{D}}

\newcommand{\md}{{\dot\mu}}
\newcommand{\nd}{{\dot\nu}}
\newcommand{\ld}{{\dot\lambda}}

\begin{document}

\begin{titlepage}
%\catcode`\@=11
%\catcode`\@=12
%\twocolumn[\hsize\textwidth\columnwidth\hsize\csname%
%@twocolumnfalse\endcsname

%\draft
\begin{center}
%\hfill hep-th/yymmnnn\\
\hfill UT-08-11
\vskip .5in

\textbf{\LARGE M2 to D2 revisited}

\vskip .5in
{\large
Pei-Ming Ho$^\dagger$\footnote{
e-mail address: pmho@phys.ntu.edu.tw}, 
Yosuke Imamura$^\ddagger$\footnote{
email address: imamura@hep-th.phys.s.u-tokyo.ac.jp},
Yutaka Matsuo$^\ddagger$\footnote{
e-mail address:
 matsuo@phys.s.u-tokyo.ac.jp}}\\
\vskip 3mm
{\it\large
$^\dagger$
Department of Physics and Center for Theoretical Sciences, \\
National Taiwan University, Taipei 10617, Taiwan,
R.O.C.}\\
\vskip 3mm
{\it\large
$^\ddagger$
Department of Physics, Faculty of Science, University of Tokyo,\\
Hongo 7-3-1, Bunkyo-ku, Tokyo 113-0033, Japan\\
\noindent{ \smallskip }\\
}
\vspace{60pt}
%\maketitle
\end{center}
\begin{abstract}

We present two derivations of the multiple D2 action from the
multiple M2-brane model proposed by Bagger-Lambert 
and Gustavsson.
The first one is to start from Lie 3-algebra associated with
given (arbitrary) Lie algebra.  The Lie 3-algebra metric is not positive
definite but the zero-norm generators merely correspond to Lagrange
multipliers.  Following the work of Mukhi and Papageorgakis,
we derive D2-brane action from the model by giving a variable a 
vacuum expectation value.
The second derivation is based on the correspondence
between M2 and M5. 
We compactify one dimension and wind
M5-brane along this direction. This leads to  a
noncommutative D4 action. Multiple D2 action is then obtained 
by suitably choosing the non-commutative parameter on the two-torus.
It also implies a natural interpretation to the extra generator in
Lie 3-algebra, namely the winding of M5 world volume around $S^1$
which defines the reduction of M theory to \IIA superstring.
\end{abstract}

%\pacs{PACS numbers: 11.25.-w, 11.25.Mj, 11.25.Sq}%]
\end{titlepage}
%\begin{narrowtext}
\setcounter{footnote}{0}

\section{Introduction}
%\subsection{Multiple M2-Brane Action}
\label{sec:BL}

The multiple M2-brane model of Bagger-Lambert 
\cite{Bagger:2006sk,Bagger:2007jr,Bagger:2007vi}
and Gustavsson 
\cite{Gustavsson:2007vu, Gustavsson:2008dy} 
is defined on Lie 3-algebras \cite{Filippov}, 
which serve as the gauge symmetry algebras for 
the M2-brane world-volume theory. 
For the consistency of these symmetries, 
we need to impose the fundamental identity 
on the Lie 3-algberas. 
But it turns out that the fundamental identities 
are extremely restrictive. 
For quite some time the only known non-trivial example 
of Lie 3-algebras is the algebra ${\cal A}_4$ 
\cite{Kawamura:2003cw}
with 4 generators and $SO(4)$ symmetry, 
until many more examples were given 
in \cite{Ho:2008bn}. 
In fact, Nambu-Poisson brackets 
\cite{Nambu,qNambu,naryLie2,Jacobian,Vaisman}
can be viewed as infinite dimensional Lie 3-algebras, 
and it can be used \cite{Ho:2008nn} to construct 
an M5-brane out of infinitely many M2-branes. 

While it is easy to find Nambu-Poisson brackets 
equipped with positive definite invariant metrics, 
all finite-dimensional examples, 
except direct sums of ${\cal A}_4$ and trivial algebras, 
have the salient feature that the invariant metric 
is never positive definite. 
It was thus conjectured in \cite{Ho:2008bn} 
(see also \cite{FigueroaO'Farrill:2002xg,Bandres:2008vf})
that there exists no other finite dimensional 
Lie 3-algebras with a positive definite metric. 
This conjecture was later proved in Refs. 
\cite{Papadopoulos:2008sk,Gauntlett:2008uf}. 
\footnote{
On the other hand, 
it was suggested \cite{Gran:2008vi} 
that the BLG model is to be studied 
only at the level of equations of motion, 
which does not require the definition of an invariant metric. 
For other interesting development on the multiple M2 theory,
see for example \cite{recent}.
}

While ${\cal A}_4$ corresponds to a certain fixed 
configuration of M2-branes in an M-fold 
\cite{Mukhi,Lambert:2008et,Distler:2008mk,Gran:2008vi}, 
other Lie 3-algebras are needed for other backgrounds. 
Thus we either dismiss the BLG model, 
or we have to accept Lie 3-algebras with zero-norm 
or negative-norm generators. 
Some may worry that the existence of 
negative-norm generators 
in the Lie 3-algebra may lead to ghosts in the BLG model. 
Thus a crucial test of the BLG model is whether 
it can make sense for a Lie 3-algebra with a metric 
which is not positive definite.
Another important task is to find Lie 3-algebras 
which will lead to $U(N)$ gauge theories for arbitrary $N$, 
in order to describe the configuration of $N$ D2-branes 
when one of the spatial dimensions is compactified. 

In this paper, 
we first construct a Lie 3-algebra as an extension 
of an arbitrary Lie algebra (section \ref{sec:Lie3}). 
We show that the BLG model based on this 
new example of Lie 3-algebra is parity invariant, 
and the zero-norm generator 
corresponds to Lagrange multipliers (section \ref{sec:BL2}). 
%pmh
Remarkably, the overall coefficient of the Lagrangian 
has the scaling symmetry, 
and thus there is no free parameter in this theory. 
However, 
we also comment (section \ref{reduction}) 
that in general one can treat the 
field components corresponding to certain 
particular generators as non-dyanmical parameters 
without breaking supersymmetry or gauge symmetry. 
This new interpretation completely removes the ghost 
for our Lie 3-algebra. 
Following Mukhi and Papageorgakis \cite{Mukhi}, 
we consider the reduction of M2 to D2-branes
(section \ref{sec:M2D2}). 
There is no ghost after compactification, 
and a spatial dimension completely disappears, 
reducing the spacetime dimension from 11 to 10. 
We find that there are no higher order terms 
in the D2-brane action, 
and the translation symmetry is manifestly preserved. 

In this approach of deriving multiple D2-branes from 
M2-branes through a finite dimensional Lie 3-algebra, 
the physical meaning of the extra generators
are not very clear.  In section \ref{sec:M5}, we present the
second derivation of D2 from M2.  It is based on the construction
of M5-brane from M2 \cite{Ho:2008nn}, where the infinite
dimensional version of the Lie 3-algebra based
the Nambu-Poisson bracket on three dimensional space
was used.  It was shown that the field content of BLG theory
is mapped to those on M5-brane which include the self-dual
two-form field.  We compactify one dimension in this internal
3 dimensional manifold and wind one direction of 
M5-brane along this direction.
We compute the BL Lagrangian in this set-up and show that
it gives rise to non-commutative D4-brane action
where the non-commutativity is infinitesimal.
We show that it is possible to generalize the algebra of
Nambu-Poisson bracket by quantization to finite non-commutativity.  When the internal space is $T^2$, by
suitably choosing the non-commutativity parameter,
one may obtain $U(N)$ symmetry on the D2-brane world volume.
In this approach, there is no problem of positivity of the norm
from the beginning and it also provides a natural interpretation of
one of the extra generators as the winding mode of M5-brane worldvolume.

\section{Lie 3-algebra from Lie algebra}
\label{sec:Lie3}

For any given Lie algebra ${\cal G}$ 
\be
[T^i, T^j] = f^{ij}{}_k T^k
\ee
with structure constants $f^{ij}{}_k$ 
and Killing form $h^{ij}$, 
we can define a corresponding Lie 3-algebra as follows. 
Let the generators of the Lie 3-algebra be denoted 
$\{ T^{-1}, T^0, T^i \}$ 
($i = 1, \cdots, \mbox{dim}\;{\cal G}$), 
where $T^i$'s are one-to-one corresponding 
to the generators of the Lie algebra ${\cal G}$. 
The Nambu bracket is defined by 
\bea
&[T^{-1}, T^a, T^b] = 0,  \label{Nb1}\\
&[T^0, T^i, T^j] = f^{ij}{}_k T^k, \\
&[T^i, T^j, T^k] = f^{ijk} T^{-1}, \label{Nb3}
\eea
where $a, b = -1, 0, 1, \cdots, \mbox{dim}\;{\cal G}$,
and 
\be
f^{ijk} \equiv f^{ij}{}_l h^{lk} 
\ee
is totally anti-symmetrized. 

One can check that the Nambu bracket, 
which is by definition skew-symmetric, 
satisfies all fundamental identities, 
that is, for all $a, b, c, d, e$,  
\be
[T^a, T^b, [T^c, T^d, T^e]] 
= [[T^a, T^b, T^c], T^d, T^e] 
+ [T^c, [T^a, T^b, T^d], T^e] 
+ [T^c, T^d, [T^a, T^b, T^e]]. 
\ee

The requirement of invariance of the metric 
\be
\la [T^a, T^b, T^c], T^d \ra 
+ \la [T^c, [T^a, T^b, T^d] \ra = 0 
\ee
implies that the metric has to be defined as 
\bea
&\la T^{-1}, T^{-1} \ra = 0, 
\qquad 
\la T^{-1}, T^0 \ra  = -1, 
\qquad
\la T^{-1}, T^i \ra = 0, \\ 
& \la T^0, T^0 \ra = K, 
\qquad 
\la T^0, T^i \ra = 0, \\ 
&\la T^i, T^j \ra = h^{ij}, 
\eea 
where $K$ is an arbitrary constant and 
$i, j = 1, \cdots, \mbox{dim}\;{\cal G}$. 

Note that there is an algebra homomorphism
\be
T^0 \rightarrow T^0 + \a T^{-1}, 
\ee
that preserves the 3-algebra, 
but changes the metric by a shift of $K$:
\be
K = \la T^0, T^0 \ra \rightarrow 
K - 2\a. 
\ee
Thus one can always choose $T^0$ 
such that 
\be
K = 0. 
\ee

This Lie 3-algebra has the following interesting 
properties. 
\begin{enumerate}
\item
The Lie 3-algebra reduces to the Lie algebra 
when one of the slots of the Nambu bracket 
is taken by $T^0$. 
That is, 
\be
[T^0, T^i, T^j] = [T^i, T^j], 
\ee
where the bracket on the right hand side 
is the Lie algebra bracket. 
\item
The generator $T^0$ never appears 
on the right hand side of a Nambu bracket. 
\item
The generator $T^{-1}$ is central, 
that is, the Nambu bracket vanishes 
whenever $T^{-1}$ appears. 
\item 
There are negative-norm generators. 
The norm of $T^0 + \a T^1$ is $K - 2\a$, 
which is negative for sufficiently large $\a$. 
$T^1$ is a zero-norm generator.
%pmh
\item 
Generally speaking, the scaling of structure constants 
\be \label{scale}
f^{abc}{}_d \rightarrow g^2 f^{abc}{}_d
\ee 
defines a new Lie 3-algebra, 
since the scaled structure constants must 
also satisfy all the fundamental identities. 
We can scale the generators 
$T^a \rightarrow g \, T^a$ 
to absorb this scaling, 
so that the structure constants are scaled back 
to their original values, 
but this will result in a scaling of the metric 
$h^{ab} \rightarrow g^2 h^{ab}$. 
However, for the particular Lie 3-algebra 
under investigation, 
a scaling of the structure constants (\ref{scale}) 
can be absorbed by the scaling 
\be \label{scale2}
T^0 \rightarrow g^2 \, T^0, \quad 
T^{-1} \rightarrow g^{-2}\, T^{-1}, \quad 
T^i \rightarrow T^i, 
\ee
which does not change the metric at all. 
\end{enumerate}

These properties will be important for 
the consideration of multiple M2-branes.

\section{Bagger-Lambert Lagrangian} 
\label{sec:BL2}

In this section we apply the Lie 3-algebra 
constructed in the previous section to 
the Bagger-Lambert action
\cite{Bagger:2006sk,Bagger:2007jr,Bagger:2007vi}, 
which is a supersymmetric action proposed to 
describe multiple M2-branes: 
\be
S = T_2 \int d^3 x \; {\cal L}, 
\ee
where $T_2$ is the M2-brane tension, 
and the Lagrangian density ${\cal L}$ is 
\be\label{BLaction}
{\cal L} = -\frac{1}{2} \la D^{\mu}X^I, D_{\mu} X^I\ra 
+ \frac{i}{2} \la\bar\Psi, \Gamma^{\mu}D_{\mu}\Psi\ra 
+\frac{i}{4} \la\bar\Psi, \Gamma_{IJ} [X^I, X^J, \Psi]\ra 
-V(X) + {\cal L}_{CS}. 
\ee 
Here $D_{\mu}$ is the covariant derivative 
\be
(D_\mu  X^I(x))_a = \partial _{\mu} X^I_a -{f^{cdb}}_a A_{\mu c d}(x) X^I_b, 
\ee
$V(X)$ is the potential term defined by 
\be
V(X) = \frac{1}{12}\la [X^I, X^J, X^K], [X^I, X^J, X^K]\ra, 
\ee
and the Chern-Simons term for the gauge potential is 
\be\label{CS}
{\cal L}_{CS} = \frac{1}{2}\epsilon^{\mu\nu\lam}
\left(f^{abcd}A_{\mu ab}\del_{\nu}A_{\lam cd} 
+ \frac{2}{3} f^{cda}{}_g f^{efgb} A_{\mu ab} A_{\nu cd} A_{\lam ef} \right). 
\ee
The indices $I,J,K = 3,\cdots,10$, 
and they specify the transverse directions
of M2-branes; 
$\mu,\nu = 0,1,2$, describing the longitudinal directions.
The indices $a, b, c$ take values in 
$-1, 0, 1, \cdots,\mbox{dim}\;{\cal G}$ for our Lie 3-algebra 
introduced in the previous section.

The mode expansions of the fields are 
\bea
X^I &\equiv& X^I_a T^a = 
X^I_0 T^0 + X^I_{-1} T^{-1} + \Xh^I, \label{mode1} \\
\Psi &\equiv& \Psi_a T^a = 
\Psi_0 T^0 + \Psi_{-1} T^{-1} + \Psih, \\
A_{\mu} &\equiv& A_{\mu a b} T^a \otimes T^b \nn \\
&=&
T^{-1} \otimes A_{\mu(-1)} - A_{\mu(-1)} \otimes T^{-1} + 
% 0520
%\Ah_{\mu} + A'_{\mu}, 
T^0\otimes\Ah_{\mu} - \Ah_{\mu}\otimes T^0 
+ A_{\mu ij} T^i \otimes T^j, 
\label{mode3}
\eea
where 
\bea 
&\Xh \equiv X_i T^i, \qquad \Psih \equiv \Psi_i T^i, \\
&A_{\mu(-1)} \equiv A_{\mu (-1) a} T^a, 
\qquad
\Ah_{\mu} \equiv 2 A_{\mu 0i} T^i. 
\eea
% 0520
We also define 
\be
A'_{\mu} \equiv A_{\mu ij} f^{ij}{}_k T^k. 
\ee
We will see below that $A_{\mu(-1)}$ are completely 
decoupled in the BLG model, 
and $X_{-1}^I$ and $\Psi_{-1}$ are Lagrange multipliers. 

The action has $N=8$ maximal SUSY in $d=3$,
and the SUSY transformations are 
\bea
\d X^I_a &=& i\bar{\eps}\Gamma^I \Psi_a, \\
\d \Psi_a &=& D_{\mu}X^I_a \Gamma^\mu\Gamma^I \eps  
- \frac{1}{6} X^I_b X^J_c X^K_d f^{bcd}{}_a \Gamma^{IJK}\eps, \\
\d \tilde{A}_{\mu}{}^b{}_a &=& 
i\bar{\eps}\Gamma_{\mu}\Gamma_I X^I_c \Psi_d f^{cdb}{}_a, 
\qquad 
\tilde{A}_{\mu}{}^b{}_a \equiv A_{\mu cd} f^{cdb}{}_a.  
\eea
In terms of the modes, we have 
\bea
\d X^I_0 &=& i\bar{\eps} \Gamma^I \Psi_0, \label{susyfst}\\ 
\d X^I_{-1} &=& i\bar{\eps} \Gamma^I \Psi_{-1}, \\
\d \Xh^I &=& i\bar{\eps} \Gamma^I \Psih, \\
\d \Psi_0 &=& \del_{\mu} X^I_0 \Gamma^{\mu}\Gamma^I \eps, \\
\d \Psi_{-1} &=& (\del_{\mu} X^I_{-1} - \la A'_{\mu} X^I \ra ) 
\Gamma^{\mu}\Gamma^I\eps
-\frac{1}{3} \la \Xh^I \Xh^J \Xh^K \ra \Gamma^{IJK} \eps, \\
\d \Psih &=& \Dh_{\mu}\Xh^I\Gamma^{\mu}\Gamma^I\eps
- \frac{1}{2}X_0^I [\Xh^J, \Xh^K] \Gamma^{IJK} \eps, \\
\d \Ah_{\mu} &=& i\bar{\eps}\Gamma_{\mu}\Gamma_I
(X_0^I\Psih - \Xh^I \Psi_0), \\
\d A'_{\mu} &=& i\bar{\eps} \Gamma_{\mu}\Gamma_I 
[\Xh^I, \Psih]. \label{susylst}
\eea 

The gauge symmetry for the bosonic fields are written as,
\ba
\delta X^I_a= \Lambda_{cd} {f^{cdb}}_{a} X^I_b\,,\quad
\delta \tilde A_\mu{}^b{}_a=\partial_\mu \tilde\Lambda^b{}_a-
\tilde \Lambda^b{}_c \tilde A_\mu{}^c{}_a +\tilde A_\mu{}^b{}_c
\tilde \Lambda^c{}_a\,. \label{gaugetransf}
\ea
(The gauge transformation of $\Psi$ is the same as $X^I$.) 
In terms of the mode expansions, they are 
\bea
\d X^I_0 &=& 0, \label{gaugefst}\\
\d X^I_{-1} &=& \la \Lam' ,\Xh^I \ra, \\
\d \Xh^I &=& [ \hat{\Lam}, \Xh^I ], \\ 
\d \Ah_{\mu} &=& \del_{\mu} \hat{\Lam} 
- [\Ah_{\mu}, \hat{\Lam}], \\
\d A'_{\mu} &=& \del_{\mu} \Lam' - [\Ah_{\mu}, \Lam']
- [ A'_{\mu}, \hat{\Lam} ],  \label{gaugelst}
\eea
where 
\be
\hat{\Lam} = 2\Lam_{0i} T^i, \qquad 
\Lam' = \Lam_{ij} f^{ij}{}_k T^k.
\ee

Plugging the mode expansions (\ref{mode1}-\ref{mode3}) 
into the Lagrangian (\ref{BLaction}),
we get,  
up to total derivatives, 
\bea 
\mathcal{L} &=& \left\la - \frac{1}{2} (\Dh_{\mu}\Xh^I - A'_{\mu} X^I_0)^2 
+ \frac{i}{4} \bar{\Psih} \Gamma^{\mu} \Dh_{\mu}\Psih 
+ \frac{i}{4} \bar{\Psi}_0 \Gamma^{\mu} A'_{\mu}\Psih
+ \frac{1}{4} (X_0^K)^2 [\Xh^I, \Xh^J]^2 \right. \nn \\
&& \left. - \frac{1}{2} (X_0^I[\Xh^I, \Xh^J])^2 
+ \frac{1}{2} \eps^{\mu\nu\lam} \Fh_{\mu\nu} A'_{\lam} 
\right\ra
% + L_0 + L_{-1}
 + \mathcal{L}_{\rm gh}
 , \label{L1}
\eea
where 
%\bea
%L_0 &\equiv& - \left\la
%\del_{\mu} X^I_0 A'_{\mu} \Xh^I 
%\right\ra, \\ 
%L_{-1} &\equiv& \left\la
%(\del_{\mu} X^I_0)(\del_{\mu} X^I_{-1}) 
%- \frac{i}{2} \bar{\Psi}_{-1} \Gamma^{\mu}\del_{\mu} \Psi_0 \right\ra, 
%\eea
\bea
\mathcal{L}_{\rm gh} &\equiv& - \left\la
\del_{\mu} X^I_0 A'_{\mu} \Xh^I 
+
(\del_{\mu} X^I_0)(\del_{\mu} X^I_{-1}) 
- \frac{i}{2} \bar{\Psi}_{-1} \Gamma^{\mu}\del_{\mu} \Psi_0 \right\ra, 
\eea
and 
\be 
\Dh_{\mu} X^I \equiv \del_{\mu} \Xh^I - [\Ah_{\mu}, \Xh^I], 
\qquad 
\Dh_{\mu} \Psi \equiv \del_{\mu} \Psih - [\Ah_{\mu}, \Psih], 
\qquad 
\Fh_{\mu\nu} \equiv \del_{\mu}\Ah_{\nu} - \del_{\nu}\Ah_{\mu} 
- [\Ah_{\mu}, \Ah_{\nu}]. 
\ee  

This Lagrangian is invariant under 
the parity transformation 
\bea
&x^{\mu} \rightarrow - x^{\mu}, \qquad 
\Gamma^{\mu} \rightarrow - \Gamma^{\mu}, \\
&\Xh^I \rightarrow \Xh^I, \qquad 
X_0^I \rightarrow - X_0^I, \qquad 
X_{-1}^I \rightarrow - X_{-1}^I, \\
&\Psih \rightarrow \Psih, \qquad
\Psi_0 \rightarrow - \Psi_0, \qquad 
\Psi_{-1} \rightarrow - \Psi_{-1}, \\ 
&\Ah_{\mu} \rightarrow -\Ah_{\mu}, \qquad 
A'_{\mu} \rightarrow A'_{\mu}. 
\eea

%pmh
Another symmetry of this model is the scaling transformation 
of the overall coefficient of the Lagrangian. 
Usually a scaling of the structure constants 
is equivalent to a scaling of the overall constant factor 
of the action through a scaling of all fields. 
This overall factor is then an unfixed coupling,  
which is undesirable in M theory. 
However, the situation is different for our new algebra. 
As we commented in the previous section, 
the scaling of structure constants for the new algebra 
can be absorbed by a scaling of $T^0$ and $T^{-1}$ 
without changing the metric. 
In other words, the scaling of the overall coefficient 
of the Lagrangian is a symmetry. 
Explicitly, scaling (\ref{L1}) by an overall coefficient $1/g^2$ 
can be absorbed by the field redefinition 
\bea
&\hat{X}^I \rightarrow g \hat{X}^I, \quad 
X^I_0 \rightarrow g^{-1} X^I_0, \quad
X^I_{-1} \rightarrow g^3 X^I_{-1}, \\
&\hat{\Psi} \rightarrow g \hat{\Psi}, \quad
\Psi_0 \rightarrow g^{-1} \Psi_0, \quad
\Psi_{-1} \rightarrow g^3 \Psi_{-1}, \\
&\hat{A}_{\mu} \rightarrow \hat{A}_{\mu}, \quad
A'_{\mu} \rightarrow g^2 A'_{\mu}. 
\eea
Hence this Lagrangian has no free parameter at all!

Note also that $X_{-1}^I$ and $\Psi_{-1}$ 
appear only linearly in $L_{-1}$, 
and thus they are Lagrange multipliers. 
%-----------------------------------
%
%[added]
Their equations of motion are
\be \label{constraint1}
\del^2 X^I_0 = 0,
\qquad 
\Gamma^{\mu}\del_{\mu} \Psi_0 = 0. 
\ee
Hence $X^I_0$ and $\Psi_0$ become classical fields,
in the sense that off-shell fluctuations are excluded 
from the path integral.  
Actually we can set 
\begin{equation} 
X_0^I=\mbox{constant},
\quad
\Psi_0=0,
\label{vev}
\end{equation}
without breaking the supersymmetry
(\ref{susyfst})-(\ref{susylst})
nor gauge symmetry
(\ref{susyfst})-(\ref{susylst}).

After we set (\ref{vev}), 
the Lagrangian is
given by (\ref{L1})
without the last term $L_{\rm gh}$.
It is remarkable that
the ghost degrees of freedom 
associated with $X_{-1}^I$ and $\Psi_{-1}$ 
have totally disappeared for this background. 
The resulting theory is clearly 
a well defined field theory without ghosts. 

The fact that the background (\ref{vev}) 
does not break any symmetry 
suggests an alternative viewpoint towards 
the BLG model. 
That is, we can change the definition of the BLG model 
by defining $X^I_0$, $\Psi_0$ as
non-dynamical constant parameters fixed by (\ref{vev}). 
The resulting model has as large symmetry 
as the original definition of the BLG model, 
but has no ghosts. 
In this interpretation, the parameter $X^I_0$ 
plays the role of coupling constant.

%[added]
%
%-----------------------------------
%
%[removed]
%
%Consider the quantization of the theory 
%in terms of the path integral formulation. 
%We first integrate out $X^I_{-1}$ and $\Psi_{-1}$. 
%This gives 
%the constraints due to the Lagrange multipliers 
%\be \label{xxxconstraint1}
%\del^2 X^I_0 = 0,
%\qquad 
%\Gamma^{\mu}\del_{\mu} \Psi_0 = 0. 
%\ee
%Hence $X^I_0$ and $\Psi_0$ become classical fields 
%in the sense that their off-shell configurations are 
%removed from the path integral 
%\be
%{\cal Z} = \int DX_0 D\Psi_0 \d(\del^2 X_0) 
%\d(\Gamma^{\mu}\del_{\mu}\Psi_0) {\cal \hat{Z}}(X_0, \Psi_0),  
%\ee
%and we naturally shift our attention to 
%the path integral with fixed classical configurations 
%of $X_0$ and $\Psi_0$
%\be \label{xxxZp}
%{\cal \hat{Z}}(X_0, \Psi_0) = \int D\Xh D\Psih D\Ah DA' 
%e^{-\int d^3 x \la {\cal L}'(X_0, \Psi_0) \ra}, 
%\ee
%where the Lagrangian is now 
%given by (\ref{L1}) without the last two terms $L_0$, $L_{-1}$.
%We interpret ${\cal \hat{Z}}(X_0, \Psi_0)$ 
%as the partition function for a given classical background 
%$(X_0, \Psi_0)$. 
%
%
%[removed]
%
%---------------------------

\section{Reduction of 3-algebras in BLG model}
\label{reduction}

From the example of the new 3-algebra described above, 
we see that in general there are two kinds of 3-algebra generators 
that are special from the viewpoint of the BLG model.

First, if a generator $T^A$ can never be generated 
through a Nambu bracket (like $T^0$ in our 3-algebra), i.e.
\be
f^{abc}{}_A = 0 \quad \forall a, b, c, 
\ee
% 0520 
then $\tilde{A}_{\mu}{}^b{}_A = 0$, and 
it is straightforward to check that for 
the assignment 
\be \label{assign}
X^I_A = \mbox{constant}, \quad
\Psi_A = 0
% 0520
%, \quad 
%A_{\mu}{}^a{}_A = 0 
\ee
on the components corresponding 
to this generator $T^A$, we have 
$D_{\mu} X^I_A = 0$ and
the SUSY transformations of the fixed components vanish 
\be
\d X^I_A = \d \Psi_A = 0
% 0520 
% = \d \tilde{A}_{\mu}{}^b{}_A = 0
\ee
for arbitrary SUSY transformation parameter $\eps$. 
Thus the complete SUSY is preserved by (\ref{assign}). 

For the gauge symmetry, 
if we define the gauge transformation parameter 
in (\ref{gaugetransf}) as 
\be
\tilde{\Lam}^b{}_a = \Lam_{cd} f^{cdb}{}_a, 
\ee
then for arbitrary $\Lam_{cd}$, 
we have all gauge transformations of the fixed components 
vanish. 
%\be
%\d X^I_A = \d \Psi_A = \d \tilde{A}_{\mu}{}^b{}_A = 0.
%\ee
Hence the gauge symmetry is preserved for arbitrary $\Lam_{cd}$. 
However, there is the possibility that in some cases 
not all degrees of freedom in $\tilde{\Lam}^b{}_a$ 
correspond to $\Lam_{cd}$, 
and the corresponding gauge symmetry may be broken, 
while all those which can be written in terms of $\Lam_{cd}$ are preserved.

Similarly, if a generator $T^A$ is central (like $T^{-1}$ in our 3-algebra), i.e., 
\be
f^{Aab}{}_c = 0 \quad \forall a, b, c, 
\ee
then the assignment 
\be \label{assign2}
X^I{}^A = \mbox{constant}, \quad 
\Psi^A = 0 
% 0520
%\quad A_{\mu}{}^{bA} = 0
\ee
preserves SUSY and gauge symmetry. 
Here the index $A$ is raised using the invariant metric 
\be
X^I{}^A \equiv X^I_a \, h^{aA}, \quad \mbox{etc}.
\ee

Furthermore, corresponding to the central element $T^A$, 
the components 
\be \label{centro-comp}
X^I_A, \quad \Psi_A, \quad \tilde{A}_{\mu}{}^b{}_A 
\ee
cannot appear in the interaction terms. 
$X^I_A$ and $\Psi_A$ can only appear in the kinetic terms, 
while $\tilde{A}_{\mu}{}^b_A$ is completely decoupled. 

Since the metric components for central elements are 
not constrained by the requirement of invariance, 
we can always choose them to vanish
\be
h^{AB} = 0,
\ee 
and the components $X^I_A$ and $\Psi_A$ can 
only appear linearly in the kinetic terms. 
They can then be integrated out as Lagrange multipliers.

As the assignments (\ref{assign}) and (\ref{assign2}) 
for two special types of generators 
preserve all SUSY and gauge symmetries, 
one can take the viewpoint that these variables 
are non-dynamical {\em by definition}. 
We have seen earlier that this interpretation 
removes the ghost from the BLG model for 
our new 3-algebra.

\section{From M2 to D2} 
\label{sec:M2D2}

Let us now consider the theory
% 0520
defined in section 3 
for the particular background 
\be \label{v}
X^I_0 = v^I, \qquad \Psi_0 = 0, 
\ee
where $v$ is a constant vector. 
Without loss of generality, 
% 0520
for space-like vector $v$, 
we can choose $v$ to lie on the direction of $X^{10}$
\be
v^I = v \, \d^I_{10}. 
\ee
As we mentioned in the previous section,
fixing the fields $X_0^I$ and $\Psi_0$ by (\ref{v})
removes the ghost term $\mathcal{L}_{\rm gh}$ from the
Lagrangian.
We can now integrate over $A'$ and find 
%
%----------------
%
%[removed]
%
%\be
%{\cal \hat{Z}}(v, 0) = \int D\Xh D\Psih D\Ah 
%e^{-\int d^3 x \la L_{\mbox{\tiny eff}} \ra }, 
%\ee
%where 
%
%[removed]
%
%-------------------------
%
\be \label{D2}
\mathcal{L}_{\mbox{\small eff}} = 
-\frac{1}{2}(\Dh_{\mu} \Xh^A)^2 + 
\frac{1}{4} v^2 [\Xh^A, \Xh^B]^2 
+ \frac{i}{4} \bar{\Psih} \Gamma^{\mu}\Dh_{\mu} \Psih 
- \frac{1}{4 v^2} \Fh^2_{\mu\nu}, 
\ee
where $A, B = 3, \cdots, 9$.

It is very interesting to note that 
all degrees of freedom in the spatial coordinate $X^{10}$ have 
totally disappeared from both the kinetic term 
and the potential term of the action. 
It is fully decoupled from the Lagrangian 
for the particular background under consideration.

Let us now recall that when M theory is compactified 
on a circle, 
it is equivalent to type \IIA superstring theory 
and M2-branes are matched with D2-branes. 
The background (\ref{v}) considered above 
is reminiscent of the novel Higgs mechanism in \cite{Mukhi}.
It was originally proposed to describe 
the effect of compactification of $X^{10}$, 
and later found to correspond to a large $k$ 
limit of a $\mathbb{Z}_{2k}$ M-fold 
\cite{Lambert:2008et,Distler:2008mk}.

%If we use the convention that the
%components of the metric $h^{ij}$ for the Lie algebra are
%numbers which does not depend on any parameters
%in the theory,
%we can identify the parameter $v$ with
%the Yang-Mills coupling constant $g_{\rm YM}\sim g_{\rm str}^{1/2}$.
%By rescaling the scalar fields
%as $\hat X^A\rightarrow \hat X^A/g_{\rm YM}$,
%we obtain the ordinary D2-brane action
%with overall factor $T_{\rm D2}\sim 1/g_{\rm str}$.

The M theory parameters can be converted to 
the parameters of type \IIA superstring theory via 
\be
R = g_s l_s, \qquad \mbox{and} \qquad 
T_s \equiv \frac{1}{2\pi \a'} = 2\pi R T_2.
\ee
The Lagrangian (\ref{D2}) is thus 
exactly the same as the low energy effective action 
of multiple D2-branes if $v$ is given by 
the perimeter of the compactified dimension
\be
v = 2\pi R. 
\ee

Despite the similarity, there are a few features 
of our model that are different from \cite{Mukhi}:
\begin{enumerate}
\item
The action (\ref{D2}) does not have higher order terms. 
\item 
The translation symmetry of the center of mass coordinates 
corresponding to the $u(1)$ factor of ${\cal G}$ is manifest. 
\end{enumerate}
These are considered as stronger signatures of 
the reduction of M2 to D2 due to a compactification 
of the M theory on $S^1$.

% 0520 
As the D2-brane is dual to M2-brane, 
the 11-th dimension of the M theory 
is not lost when $X^{10}$ disappears. 
It is dual to the gauge field degrees of freedom 
on the D2-brane \cite{Duff:1992hu}.

\section{From M5 to D2}
\label{sec:M5}

In this section, we present a very different derivation of D2-brane 
from M2.  It is based on the derivation of M5-brane from BLG
theory \cite{Ho:2008nn}.  We consider a three dimensional
manifold $\cN$ equipped with the 
Nambu-Poisson structure.  By choosing the appropriate local 
coordinates $y^\md$ ($\md=\dot 1,\dot2,\dot3$),
one may construct an infinite dimensional Lie 3-algebra
from the basis of functions on $\cN$, $\chi^a$ ($a=1,2,3,\cdots$)
as,
\ba
\left\{ \chi^a, \chi^b, \chi^c\right\}
=\sum_d {f^{abc}}_d \,\chi^d\,,\quad
\left\{f_1, f_2, f_3\right\}
=\sum_{\md,\nd,\ld} \epsilon_{\md\nd\ld}
\frac{\partial f_1}{\partial y^\md}
\frac{\partial f_2}{\partial y^\nd}
\frac{\partial f_3}{\partial y^\ld}\,.
\ea
{}From the property of the Nambu-Poisson structure,
this 3-algebra satisfies the fundamental identity
with positive definite and invariant metric for the generators,
\ba
\langle\chi^a, \chi^b\rangle =\int _\cN d^3 y \chi^a(y) \chi^b(y)\,.
\ea
By the summation of these generators with the fields in
BL action,
\ba
&& X^I(x,y)= \sum_{a} X^I_a(x) \chi^a(y)\,,\\
&& \Psi(x,y)=\sum_a \Psi_a(x)\chi^a(y)\,,\\
&& A_\mu(x,y,y')=\sum_{a,b} A_{\mu a b}(x)\chi^a(y) \chi^b(y'), 
\ea
we obtain the fields on the six dimensional manifold
$\cM\times\cN$ where $\cM$ is the world volume of the original 
membrane. We note that the gauge field $A_\mu(x,y,y')$ appears
to depends on two points on $\cN$.  However, if we examine the
action carefully, one can show that it depends on $A_\mu(x,y,y')$
only through \cite{toappear},
\ba
b_{\mu \nd}(x,y)=\left.\frac{\partial}{\partial y'^\nd} A_\mu(x,y,y')\right|_{y'=y}\,.
\ea
Therefore the action can be written in terms of the local fields.
It was shown that the BL Lagrangian, after suitable field redefinitions,
describes the field theory on M5 \cite{Ho:2008nn}
which includes the self-dual two-form field.
While the analysis in \cite{Ho:2008nn} is at the level of quadratic order,
we will present here the nonlinear action which includes all the terms
in BL action.  This is based on a technical development
in \cite{toappear} where the exact analysis including the nonlinear
terms are given.  Because the full detail of the computation is
given in \cite{toappear}, we present only the result and its implication here.

In order to obtain D4 from M2, we have to wind
$X^{\dot 3}$ around
the compact $y^{\dot 3}$ direction \cite{r:M5D4}
and impose the constraints
that the other fields do not depend
on $y^{\dot 3}$.  Other than that, we use the same field configuration
\cite{Ho:2008nn}
\ba
&&X^{\dot 3}= y^{\dot 3}\,,\\
&& X^{\dot \alpha} = y^{\dot\alpha}+\epsilon_{\dot\alpha\dot\beta}
a_{\dot\beta}(x,y)\,,\\
&& a_\mu(x,y)=b_{\mu \dot 3}(x,y)\,,\\
&& \tilde a_\lambda(x,y)=\epsilon_{\dot\alpha\dot\beta}
\partial_{\dot\alpha} b_{\lambda\dot\beta}\,,\\
&& \partial_{\dot 3} X^i=\partial_{\dot 3}\Psi=
\partial_{\dot 3}a_{\dot\beta}
=\partial_{\dot 3}a_\mu=\partial_{\dot 3}\tilde a_\lambda=0\,.
\ea
Here we use the indices $\dot\alpha,\dot \beta,\cdots$ to denote
$\dot 1, \dot 2$ such that the world volume index of D4 is
$\mu$ and $\dot\alpha$.  We use the notation $i=1,\cdots,5$
for the transverse directions.
We repeat the same computation as in \cite{Ho:2008nn}
but here we include the nonlinear terms.
It turns out that $b_{\mu\nd}$ appears only through $a_\mu$
and $\tilde a_\mu$.  

Various terms of the D4 action can be computed \cite{Ho:2008nn,toappear}
straightforwardly.  First the potential term becomes 
\ba
&& -\frac{1}{12}\la [X^I, X^J, X^K]^2\ra\nn\\
&&~~~~~~~~~
=\int_\cN d^3 y\left( -\frac{1}{2}-F_{\dot1\dot2}-\frac{1}{4} {F_{\dot\alpha \dot\beta}}^2
-\frac{1}{4}{{\cD}_{\dot\alpha} X_i}^2
-\frac{1}{4}\left\{
X_i, X_j
\right\}^2\right),
\ea
where
\ba
F_{\dot\alpha\dot\beta}:=\partial_{\dot\alpha} a_{\dot \beta}
-\partial_{\dot\beta} a_{\dot \alpha}+\left\{
a_{\dot \alpha},a_{\dot \beta}
\right\}
\,,\quad
\cD_{\dot\alpha} X_i=\partial_{\dot\alpha} X_i +\left\{
a_{\dot\alpha}, X_i
\right\}\,.
\ea
While we expect to have the Abelian $U(1)$ gauge field on the world
volume,  we have the Poisson bracket
\ba
\left\{ f,g\right\}=\sum_{\dot\alpha,
\dot\beta=\dot 1,\dot 2} \epsilon_{\dot\alpha \dot\beta}
\partial_{\dot\alpha} f \partial_{\dot\beta} g
\ea
 everywhere.
It implies that we can not escape from the noncommutativity
in $\cN$ direction as long as we start from BL Lagrangian.
The Chern-Simons term (\ref{CS}) becomes,
after partial integrations,
\ba
&& {\cal L}_{CS} =-\frac{1}{2}\epsilon^{\mu\nu\lambda}
\int d^3 y \,\tilde a_\mu(x,y) F_{\nu\lambda}(x,y)\,,\quad
F_{\mu\nu}:=\partial_\mu a_\nu -\partial_\nu a_\mu
+\left\{
a_\mu, a_\nu
\right\}\,.
\ea
Finally the kinetic terms for $X^I$ and the fermion become
\ba
&&-\frac{1}{2}\la {(D_\mu X^I)^2}\ra=-\frac{1}{2} 
\int_\cN d^3 y\left({F_{\mu\dot\alpha}}^2
+{\tilde a_\mu}^2+{\cD_\mu X_i}^2\right)\,,
\\
&& \frac{i}{2} \la \bar\Psi,\Gamma^\mu D_\mu \Psi\ra
+\frac{i}{4}\la \bar\Psi,\Gamma_{IJ}[X^I, X^J, \Psi]\ra\\
&& ~~~=\frac{i}{2}\int_\cN d^3 y \left(\
\bar\Psi\Gamma^\mu\cD_\mu\Psi
+\bar\Psi\Gamma^{\dot \alpha} \cD_{\dot\alpha}\Psi
+\bar\Psi \Gamma_i\left\{ X^i,\Psi\right\}
\right)\,,
\ea
where
\ba
&&
F_{\mu\dot\alpha}:=\partial_\mu a_{\dot\alpha}-
\partial_{\dot\alpha} a_\mu+\left\{
a_\mu, a_{\dot\alpha}
\right\}\,,\quad
\cD_\mu X_i=\partial_\mu +\left\{ a_\mu, X_i\right\}\,,\\
&& \cD_\mu \Psi=\partial_\mu\Psi+\left\{a_\mu, \Psi\right\},\quad
\cD_{\dot\alpha}\Psi=\partial_{\dot\alpha}\Psi+\left\{a_{\dot\alpha},
\Psi\right\}\,,\\
&& \Gamma_{\dot\alpha}=\sum_{\dot\beta}\Gamma_{\dot 3\dot\beta}\epsilon_{\dot\beta
\dot\alpha}\,,\quad
\Gamma_i=\Gamma_{\dot 3 i}\,.
\ea
We note that the field $\tilde a_\mu$ does not have the kinetic term
and can be integrated out exactly. 
The integrand does not depend on $y^{\dot 3}$ so we obtain
overall factor of $2\pi R$ ($R$ is the radius of the compactified direction)
after the integration over $y^{\dot 3}$.

We note that in the computation, there are no ambiguities associated
with the inner product.  After integrating out the auxiliary field
$\tilde a_\mu$, one arrives at the D4-brane action
(after neglecting the constant term and the total derivative term)
\ba
S=2\pi R\int d^5 x \left(
-\frac{1}{4} {F_{\mh \nh}}^2-\frac{1}{2}{\cD_{\mh} X^i}^2+
\frac{i}{2} \bar\Psi \Gamma^{\mh}\cD_{\mh} \Psi
-\frac{1}{4}{\left\{X^i,X^j\right\}}^2
+\frac{i}{2} \bar\Psi \Gamma_i\left\{ X^i,\Psi\right\}
\right)\,.
\ea
Here $\mh,\nh,\cdots$ are the integrated indices for $\mu,\nu$ and
$\dot\alpha,\dot\beta$ run from $0$ to $4$.  As already mentioned,
$A_{\mh} =a_\mu, a_{\dot\alpha}$ is not exactly the commutative
U(1) gauge field but it includes noncommutativity in $\mh=3,4$ directions
(originally $\dot\alpha$ directions). 
The definition of the field strength and the covariant derivatives
are, of course,
\ba
&&F_{\mh\nh}=\partial_{\mh}A_{\nh}-\partial_{\nh}A_{\mh}+\left\{
A_{\mh},A_{\nh}\right\}\,,\nn\\
&&\cD_{\mh} X^i=\partial_{\mh} X^i +\left\{A_{\mh}, X^i\right\}\,,\quad
\cD_{\mh} \Psi=\partial_{\mh} X^i +\left\{A_{\mh}, \Psi\right\}\,.
\ea

The origin of the noncommutativity is obvious.
It comes from the Nambu-Poisson bracket where
the space of the function is truncated to
\ba\label{restricted}
\left\{y^3\right\}\cup C(\mathcal{N}')\,.
\ea
Here we decompose $\cN$ into $y^{\dot 3}$ direction
and $\cN'$ described by $y^{\dot1,\dot2}$.
The Nambu-Poisson bracket becomes 
(for $f_i(y^{\dot1}, y^{\dot2}) \in C(\mathcal{N}')$) 
\ba
\left\{y^3, f_1, f_2\right\}^{NP}=\left\{f_1, f_2\right\}\,,\quad
\left\{ f_1, f_2, f_3\right\}^{NP}=0\,,\quad
 \mbox{others}=0\,.
\ea
The commutator terms in the lagrangian come from
this algebra. This algebra turns out to be identical to
Lie 3-algebra (\ref{Nb1}--\ref{Nb3}) if we put $T^{-1}$ to zero.
The generator that corresponds to $T^0$ is $y^{\dot 3}$, which
describes the winding of M5 world volume around $S^1$.

The Poisson bracket $\left\{f,g\right\}$ can be obtained
from the matrix algebra when the matrix size $N$ is infinite.
By using the standard argument (see for example \cite{rD2D4}), 
it is easy to claim that
the D4 action which we just obtained can be regarded as
describing an infinite number of D2-branes.

However, in order to obtain the finite $N$ theory on D2-brane,
this is not sufficient. We need to quantize the Nambu
bracket.  In general, the quantum Nambu bracket is
very difficult to define.  However,
for the truncated Hilbert space (\ref{restricted}),
this is actually possible. We deform 
the Nambu-Poisson bracket by,
\ba
[f_1, f_2, f_3]^{QN}=\sum_{i,j,k=1}^3
\epsilon_{ijk} (f_i \star f_j) \partial_3 f_k
\label{qN}
\ea
where $\star$ is the Moyal product,
\ba
(f\star g) (y^{\dot1}, y^{\dot2})=\exp(i\epsilon_{\dot\alpha\dot\beta}
\theta \partial_{y^{\dot\alpha}}\partial_{z^{\dot\beta}}) f(y^{\dot1}, y^{\dot2})
g(z^{\dot1}, z^{\dot2})|_{z=y}\,.
\ea
It does not satisfy the fundamental identity when we consider $C(\mathcal{N})$
as a whole.
If we restrict the generators to (\ref{restricted}), we can recover
the fundamental identity.
If we take $\mathcal{N}'$ as $T^2$ and quantize $\theta$ suitably,
the quantum $T^2$ reduces to the $U(N)$ algebra,
\ba
UV=VU\omega, \quad \omega^N=1\,,
\quad U^N=V^N=1\,.
\ea
In this case the quantum Nambu-Poisson bracket reduces to
the one-generator extension of $U(N)$ algebra
\ba
[T^0, T^i, T^j]= {f^{ij}}_k T^k\,,\quad
[T^i, T^j, T^k]=0\,.
\ea
The multiple
D2 action can be obtained by expanding the functions
in $y^{\dot1,\dot2}$ directions by $U,V$ and
replacing the covariant derivative $\cD_{\dot\alpha}$
by the commutators 
\ba
\cD_{\dot\alpha} \Phi\rightarrow [X_{\dot\alpha},\Phi]\,
\ea
for general $\Phi$.

In this way, by taking a path M2 $\rightarrow$ M5 $\rightarrow$
D4 $\rightarrow$ D2, one can obtain the multiple D2 theory
without touching the problem of the negative-norm state.

\section{Conclusion}

In this paper, we study two approaches to obtain
multiple D2-brane action from the BLG theory.
In the first approach, one defines Lie 3-algebra which
contains generators of a given Lie algebra.
Such an extension inevitably contains generators with 
negative norms.  We argued that by suitably choosing
such extension, one might restrict the field associated with
it to constant or zero while keeping almost all of the
symmetry of BLG theory.  Such truncation leads to the
symmetry breaking mechanism of \cite{Mukhi}
and generates the standard kinetic term for the gauge fields
on the multiple D2-brane worldvolume.

In \cite{Ho:2008bn}, we have presented many examples of
Lie 3-algebras which satisfy the fundamental identity.
The algebra which we consider here is a generalization
of one of them.  It is quite interesting to conjecture
that similar mechanism which we consider here may be applied to
other examples by restricting the fields associated with the 
null/negative norm generators to constants.  Such theories
may not describe M2 or D2 but would give a new insight into
M theory dynamics.

In the second derivation of multiple D2-brane, we found that
the extra generator has a simple physical origin, the winding
of M5-brane around $S^1$ which defines the reduction from 
M theory
to the type \IIA theory. One may provide a similar geometrical origin
to other Lie 3-algebras.

We also commented that to have finite $N$ theory from M5, we need
quantization of the Nambu-Poisson bracket. This is trivially possible
in our case for D4-branes since 
we have reduced the Namb-Poisson bracket into 
the usual Poisson bracket.
In general, however, we need to consider the quantization of
full Nambu-Poisson bracket in the full function space.
We hope that the many studies in the past 
\cite{Awata:1999dz,qNambu,Ho:2007vk} would provide a breakthrough
toward this direction.

\section*{Note added}
When we have almost 
finished the paper, there appeared a paper
\cite{Gomis:2008uv} which overlaps considerably on the first proposal
of this paper for deriving D2 from M2 in the BLG model.

\section*{Acknowledgment}

We appreciate partial financial support from
Japan-Taiwan Joint Research Program
provided by Interchange Association (Japan)
by which this collaboration is made possible.

The authors thank Kazuyuki Furuuchi, Darren Sheng-Yu Shih, 
and Wen-Yu Wen for helpful discussions. 
P.-M. H. is grateful to Anna Lee for assistance in many ways. 
The work of P.-M. H. is supported in part by
the National Science Council,
and the National Center for Theoretical Sciences, Taiwan, R.O.C. 
Y. M. is partially supported by
Grant-in-Aid (\#20540253) from the Japan
Ministry of Education, Culture, Sports,
Science and Technology.
Y.I. is partially supported by
a Grant-in-Aid for Young Scientists (B) (\#19740122) from the Japan
Ministry of Education, Culture, Sports,
Science and Technology.

\end{document}